\documentclass[a4paper, 10pt, twocolumn]{article}

\usepackage{amsfonts,amssymb,amsmath,amsthm}
\usepackage{subfigure}
\usepackage{graphicx}
\usepackage[footnotesize]{caption}
\usepackage{color}
\usepackage[utf8x]{inputenc}
\usepackage{multicol}
\usepackage{multirow} 
\usepackage{algorithm} 
\usepackage{enumitem}

\newcommand\eX{\ensuremath{\mathsf{\textbf{X}}}}

\newcommand\eS{\ensuremath{\mathsf{\textbf{S}}}}
\newcommand\eA{\ensuremath{\mathsf{\textbf{A}}}}

\newcommand\eU{\ensuremath{\mathsf{\textbf{U}}}}

\newcommand\eW{\ensuremath{\mathcal{W}}}
\newcommand\eR{\ensuremath{\mathbb{R}}}
\newcommand\eD{\ensuremath{\mathcal{D}}}
\newcommand\ent{\ensuremath{\operatorname{ent}}}
\newcommand\Ent{\ensuremath{\operatorname{\textbf{Ent}}}}

\newcommand{\argmin}[2]{\ensuremath{\underset{\substack{{#1}}}%
{\mathrm{Argmin}}\;\;#2 }}
\newcommand{\minimize}[2]{\ensuremath{\underset{\substack{{#1}}}%
{\mathrm{Min}}\;\;#2 }}
\newcommand{\approxim}[1]{\ensuremath{\mathrm{Appr}#1 }}

\newcommand{\betadiv}{\ensuremath{\operatorname{\beta-\textit{div}}}}
\newcommand{\betaddiv}{\ensuremath{\operatorname{\beta-\textbf{div}}}}

\usepackage[final]{changes}
\definechangesauthor[name={Afef}, color=magenta]{af}
\newcommand{\afa}[1]{\added[id=af]{#1}}
\newcommand{\afr}[2]{\replaced[id=af]{#1}{#2}}
\newcommand{\afd}[1]{\deleted[id=af]{#1}}

\definechangesauthor[name={Sandrine}, color=blue]{sa}
\newcommand{\saa}[1]{\added[id=sa]{#1}}
\newcommand{\sar}[2]{\replaced[id=sa]{#1}{#2}}

\usepackage[top=1.5cm, left=1.5cm, right=1.5cm, bottom=1.5cm]{geometry}

\title{Unmixing 2D HSQC NMR mixtures with $\beta$-NMF and sparsity}

\author{Afef Cherni, Sandrine Anthoine, Caroline Chaux\thanks{The authors would like to thank E. Piersanti, L. Shintu and M. Yemloul from iSm2, Aix-Marseille Univ., for their collaboration.
The project has received funding from the Excellence Initiative of Aix-Marseille University - A$*$Midex, a French ``Investissements d'Avenir'' program.}
\\
\footnotesize Aix-Marseille Univ, CNRS, Centrale Marseille, I2M, Marseille, France
}
\date{\empty} 

\renewenvironment{abstract}{\bf\small {\em\ Abstract---}}{}

\begin{document}

\maketitle
\begin{abstract}
Nuclear Magnetic Resonance (NMR) spectroscopy is an efficient technique to analyze chemical mixtures in which one acquires spectra of the chemical mixtures along one ore more dimensions. One of the important issues is to efficiently analyze the composition of the mixture, this is a classical Blind Source Separation (BSS) problem. The poor resolution of NMR spectra and their large dimension call for a tailored BSS method.
We propose in this paper a new variational formulation for BSS based on a  $\beta$-divergence data fidelity term combined with sparsity promoting regularization functions. A majorization-minimization strategy is developped to solve the problem and experiments on simulated and real 2D HSQC NMR data illustrate the interest and the effectiveness of the proposed method.
\end{abstract}
%

\section{Introduction}
\label{sec:intro}
Blind Source Separation (BSS) consists in estimating $N$ sources from $M$ mixtures (in this work we consider $M>N$) without knowing the mixing operator.
It appears in many \afr{fields}{applications} such as biology, chemistry, astronomy, telecommunications, etc.~\cite{comon2010handbook}.
\afd{In this paper, we are interested in nuclear magnetic resonance bidimensional data.}
Nuclear Magnetic Resonance (NMR) is a powerful tool used to characterize and determine properties of molecules present in a given chemical mixture.
\sar{Here, we are interested in NMR bidimensional  (2D) data, which }
 {NMR 2D data} are nonnegative and characterized by a high sparsity level presenting crowded spectra with an important spectral overlap and poor resolution (see Fig.~\ref{fig:2DNMRrslt} \afa{(a)}).
Designing a robust BSS approach tailored to the 2D NMR data would greatly help the analysis of NMR data which is currently mostly done by the chemists. 

The BSS problem in this context is a nonnegative matrix factorization (NMF) problem.
This concept introduced by Lee and Seung~\cite{lee1999learning} was exploited in different applications either based on the classical Frobenius distance~\cite{paatero1994positive, lee1999learning} or based on the $\beta$-divergence family of cost functions~\cite{fevotte2009nonnegative, fevotte2011algorithms}.
Moreover, different works showed that the Frobenius distance associated with regularization functions is an efficient framework enabling to solve the BSS problem.
Recently, in~\cite{cherni2019nmf} the Frobenius norm combined with various regularization functions was proposed and demonstrated its effectiveness to unmix complex NMR mixtures.
In this work, we propose to investigate a $\beta$-NMF approach in which a $\beta$-divergence is associated with regularization functions that favour sparsity.


\section{Methodology and algorithm}
\label{sec:BSS}

The forward model is the following. 
The $N$ sources composed of $L$ samples are stored row-wise in the matrix $\eS \in \eR^{N \times L}$. The measures are $M$ mixtures stored row-wise in the matrix $\eX \in \eR^{M \times L}$ that follow the model
\begin{equation}
\eX = \eD(\eA \eS),
\label{eq:pbinv}
\end{equation}
where $\eA \in \eR^{M \times N}$ is the mixing matrix and $\eD$ the degradation model that depends on the application.
The BSS problem is the joint estimation of $\eA$ and $\eS$ from $\eX$. 

As in various NMF approaches~\cite{lee1999learning, paatero1994positive}, we propose to solve this problem by minimizing a variational functional. For the fit-to-data term, we investigate the use of the so-called $\beta$-divergence (noted $\betaddiv$) as proposed in~\cite{kompass2007generalized, fevotte2011algorithms}. In addition, our functional contains regularization terms for $\eA$ and $\eS$ that encompass the nonnegativity of the entries of the matrices and the sparsity of the sources (rows of $\eS$) which will represent 2D NMR spectra in our experiments.
Our goal is thus to solve
\begin{equation}
\minimize{\eA, \, \eS}{ \Phi(\eA, \eS):= \betaddiv(\eX, \eA\eS)+ \lambda_{\eA} \Psi_{\eA}(\eA)+\lambda_{\eS} \Psi_{\eS}(\eS)},
\label{eq:minimize_betadiv_reg}
\end{equation}
with $\lambda_{\eA}>0$, $\lambda_{\eS}>0$ the regularization parameters, and $\betaddiv$, 
$\Psi_{\eA}$ and $\Psi_{\eS}$ defined below.

The fit-to-data term is measured using the $\beta$-divergence 
\begin{equation}
\betaddiv(\eX, \eX') = 
\sum_{m}\sum_{l} \betadiv(\eX_{m,l} | \eX'_{m,l}),
\label{eq:beta-div-mat}
\end{equation}
where $\betadiv$ is defined on $ (\eR_+)^2$ and for $\beta \in \eR \setminus \{0,1\}$ as~\cite{basu1998robust}
\begin{equation}
\betadiv(u | v)=
    \tfrac{1}{\beta (\beta-1)} \left(u^{\beta} + (\beta-1) v^{\beta} - \beta u v^{\beta-1} \right).
\label{eq:beta-div}
\end{equation}
Note that the $\beta$-divergence is also defined for $\beta=0$ or $1$ as respectively 
the Itakura-Saito divergence and the Kullback-Leibler divergence~\cite{hoffman1999probabilistic}. 
The choice of $\beta$ varies generally according to the context and the problem characteristics (e.g. type of noise). In this work, we investigate the range $\beta>2$.

The regularization functions $\Psi_{\eA}$ and $\Psi_{\eS}$ include the nonnegativity constraint $\iota_+(\eU) =0$ if $\eU_{i,j} \geq 0 \quad  \forall i,j $ and $\iota_+(\eU) =+\infty $ otherwise.
%
In addition for the sources, we enforce sparsity either with a classical $\ell_1$ norm 
as \sar{ used in Compressive Sensing~\cite{4472240} and many image processing methods, e.g.~\cite{doi:10.1137/080716542}
}
{in various applications such as  Compressive Sensing~\cite{4472240} and image restoration~\cite{doi:10.1137/080716542}}
or with the Shannon \saa{negative} entropy $\Ent$ as proposed in~\cite{afef2016proximity, cherni2017palma} as a sparsity promoting penalty in the NMR context.  We have: $\Ent(\eU) = \sum_{n,l} \ent(\eU_{n,l})$ where
\begin{equation}\ent(u)=
\begin{cases}
u \log(u) & \text{if } u > 0 \\
0 & \text{if } u=0 \\
+\infty & \text{otherwise}.\\
\end{cases} 
\end{equation} 
As a result, we propose to minimize Eq.~\eqref{eq:minimize_betadiv_reg} with
$\Psi_{\eA}=\iota_+$ and either \textbf{(i)} 
$\Psi_{\eS}=\iota_+ + ||\cdot||_1$ or \textbf{(ii)} $\Psi_{\eS}=\iota_+ + \Ent$. 
%
%
To solve this problem, we derive an alternating minimization procedure (as in \sar{dictionary learning}{ICA, SOBI}, NMF...) described by
$$\begin{array}{l}
\text{For}\;\;k = 0,1,\ldots\\
\left\lfloor
\begin{array}{l}
\eA_{k+1} = \approxim\left(\argmin {\eA}{ \betaddiv(\eX, \eA \eS_k)} + \lambda_{\eA} \Psi_{\eA}(\eA)\right)\hfill \textbf{(I)}\\
\eS_{k+1} = \approxim\left(\argmin {\eS}{ \betaddiv(\eX, \eA_{k+1} \eS) + \lambda_{\eS} \Psi_{\eS}(\eS)\right)}\hfill  \textbf{(II)} \\
\end{array}
\right.\\
\end{array}
$$
where $\approxim()$ denotes an approximation of the minimizer inside. \textbf{(I)} and \textbf{(II)}  are multiplicative update rules built using a Majorization-Minimization (MM) strategy~\cite{hunter2000rejoinder}: the functional in~\eqref{eq:minimize_betadiv_reg} is split into a convex part majorized by the Jensen-inequality and a concave part majorized by its tangent. 

We derived the following update rules for $\beta > 2$:
\begin{align}
\textbf{(I)}\ & \eA_{k+1} =  \left( \frac{ \left( \eX \odot (\eA_k \eS)^{\odot(\beta-2)} \right) \eS^T }{ (\eA_k \eS)^{\odot (\beta-1)} \eS^T } \right)_+^{\odot \frac{1}{\beta-1}} \odot \eA_k,
\label{eq:pos_updateA}\\
\textbf{(II.i)} \ 
&\eS_{k+1} =  \left( \frac{ \eA^T (\eX \odot (\eA \eS_k)^{\odot(\beta-2)}) - \lambda_{\eS}}{ \eA^T (\eA \eS_k)^{\odot (\beta-1)}} \right)_{+}^{\odot \frac{1}{\beta-1}}  \odot \eS_k ,
\label{eq:l1_update}\\
\textbf{(II.ii)}\  & \eS_{k+1} = \left( 
\frac{\gamma}{\alpha} \eW \left( \frac{\alpha}{\gamma} \exp (-\frac{\delta}{\gamma}) \right) 
\right)_+^{\odot \frac{1}{\beta-1}} \odot \eS_k ,
\label{eq:ent_update}
\end{align}
where $\odot$ denotes the Hadamard product, $(.)_+$ the projection onto the nonnegative set and $\eW$ the Lambert function~\cite{corless1996lambertw}, and
$$
\begin{array}{l}
\alpha = \eA^T (\eA \eS_k)^{\odot (\beta-1)} \odot \eS_k , \\
\gamma = \frac{\lambda_{\eS}}{\beta - 1} \, \eS_k ,\\
\delta = \lambda_{\eS} (\eS_k + \eS_k \odot \log(\eS_k) ) - \eA^T (\eX \odot (\eA \eS_k) ^{\odot (\beta-2)}) \odot \eS_k.\\
\end{array}
$$
Note that the same strategy can not be applied in all cases when $\beta \leq 2$.


\section{Experimental results}
\label{sec:app}
We process 2D Heteronuclear Single Quantum Coherence (HSQC) data where $5$ mixtures $\eX \in \eR^{5 \times 1024 \times 2048}$ and $4$ pure sources (\afa{$\eS_1$:} Limonene, \afa{$\eS_2$:} Nerol, \afa{$\eS_3$:} Terpinolene and \afa{$\eS_4$:} Caryophyllene) noted $\eS \in \eR^{4 \times 1024 \times 2048}$ are acquired on a Bruker Avance III 600 MHz spectrometer. Matrix $\eA$ is provided by the chemists who acquired the data.
The tensors are matricized ($\eX \in \eR^{5 \times 2097152}$ and $\eS \in \eR^{4 \times 2097152}$).

In the synthetic case, we use $\eA$ and $\eS$ \afa{described} above and \afa{we} simulate \afa{synthetic measures $\eX$ based on} the model in Eq.~\eqref{eq:pbinv} with $\eD$ an i.i.d. zero-mean Gaussian noise of standard deviation $\sigma=1.97 \times 10^4$. \afa{Then, we} apply our algorithm to \afr{estimate $\eA$ and $\eS$}{the obtained synthetic measures $\eX$}. 
The performances of the proposed approach are compared \sar{to using the popular Frobenius norm $(\frac{1}{2} \|\eX - \eA \eS \|_F^2$ as the data fidelity term}{to the case when the popular data fidelity term using the Frobenius norm $(\frac{1}{2} \|\eX - \eA \eS \|_F^2$) is used} (solved with a Block-Coordinate Variable Metric Forward Backward algorithm~\cite{chouzenoux2014variable} as  in~\cite{cherni2019challenges}). 
Both algorithms are initialized with a projection of the JADE~\cite{Cardoso_J_1993_J_blind_bngs} result onto the nonnegative space, and run for a maximum of $15,000$ iterations.
The stopping criterion is $(\| \hat{\eS}_{k+1} - \hat{\eS}_k \|_F / \| \hat{\eS}_k\|_F) \leq 10^{-6}$ and $(\| \hat{\eA}_{k+1} - \hat{\eA}_k \|_F / \| \hat{\eA}_k\|_F) \leq 10^{-6}$ \afa{where we denote by $\hat{\eS}$ and $\hat{\eA}$ the estimated sources and the estimated mixing matrix respectively}.
We evaluate the quality of \afd{estimated sources} \afr{$\hat{\eS}$}{$\eS$} with the SDR (Signal to Distortion Ratio), SIR (Signal to Interference Ratio) and SAR (Signal to Artefacts Ratio)~\cite{vincent2006performance} in dB and compute the Moreau-Amari index~\cite{moreau1994one} to evaluate \afd{the estimated mixing matrix} \afr{$\hat{\eA}$}{$\eA$}.
\afd{The results were obtained with Matlab R2018b running on Ubuntu 7.4.0-1.}

We ran the algorithm for several values of the hyperparameters. 
We present in Table~\ref{Tab:Sim_case} the SDR, SIR and SAR averaged over the sources and the Amari-index for different objective functions $\Phi$ based on the $\beta$-divergence ($\beta=3$) and Frobenius norm, with  $\lambda_{\eS}=0.1\sigma$ for the simulated case.
It is clear that the $\beta$-divergence improves SDR, SAR and SIR measures (the higher the better) and the Amari-index (the lower the better) for both proposed regularization functions, showing that it is an adapted choice of data fidelity term here.
However, when looking at the value for each source separately (not shown here) it seems that regularization parameter $\lambda_{\eS}$ could be adapted to each source $\eS_i$ for $i=1,...,4$.
\begin{table}[!ht]
\centering 
\renewcommand{\arraystretch}{1.05}
\setlength{\tabcolsep}{0.1cm}
\begin{small}
\begin{tabular}{|c|c|c|c|c|c|} \hline
Data fidelity term & $\Psi_{\eS}$ & SDR &  SIR & SAR & Amari-index\\ \cline{1-6}
 
\multirow{2}{*}{Squared Frobenius} 
& $\ell_1 + \iota_+$ 
&30.299 & 31.475 &39.462 & 0.0272 \\ \cline{2-6}

& $\Ent + \iota_+$ 
& 18.287 & 36.859 & 18.354 & 0.0090 \\ \cline{1-6}

\multirow{2}{*}{$\beta$-divergence}
& $\ell_1 + \iota_+$ 
& 36.531 & \textbf{40.853} & 41.255 & \textbf{0.0054} \\ \cline{2-6}
& $\Ent + \iota_+$ 
& \textbf{36.710} & 40.852 & \textbf{41.570} & \textbf{0.0054} \\ \cline{1-6}
\end{tabular}
\end{small}
\caption{Results on 2D simulated NMR data with $\lambda_{\eS}=0.1  \sigma$.}
\label{Tab:Sim_case}
\end{table}

\begin{figure}[!ht]
\centering
\includegraphics[width=7.2cm,height=3.9cm]{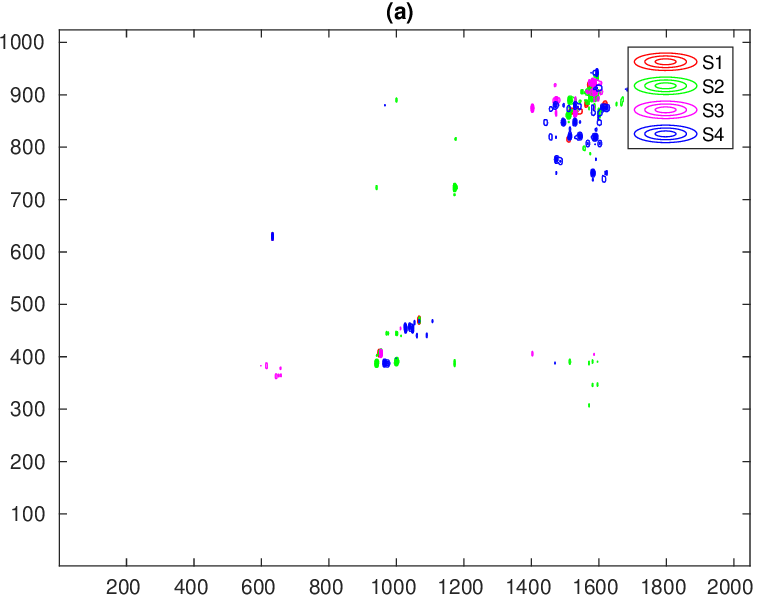}\\
\includegraphics[width=7.2cm,height=3.9cm]{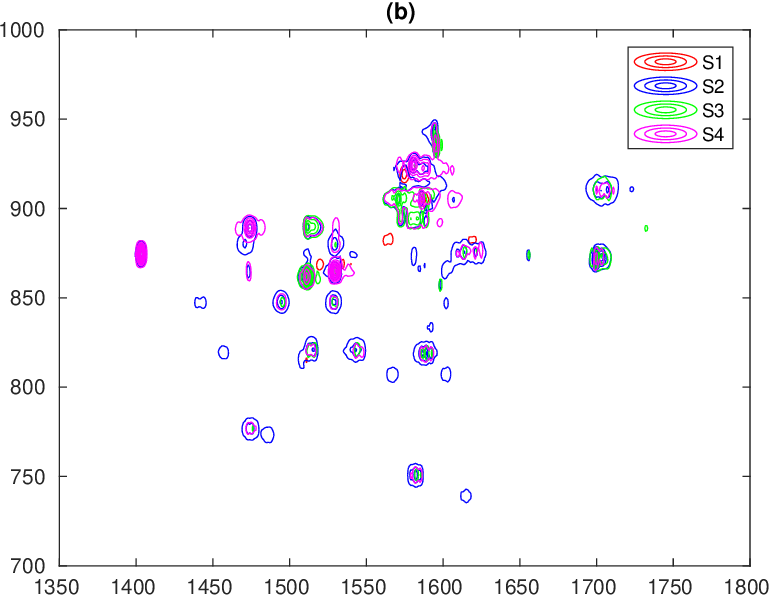}\\
\includegraphics[width=7.2cm,height=3.9cm]{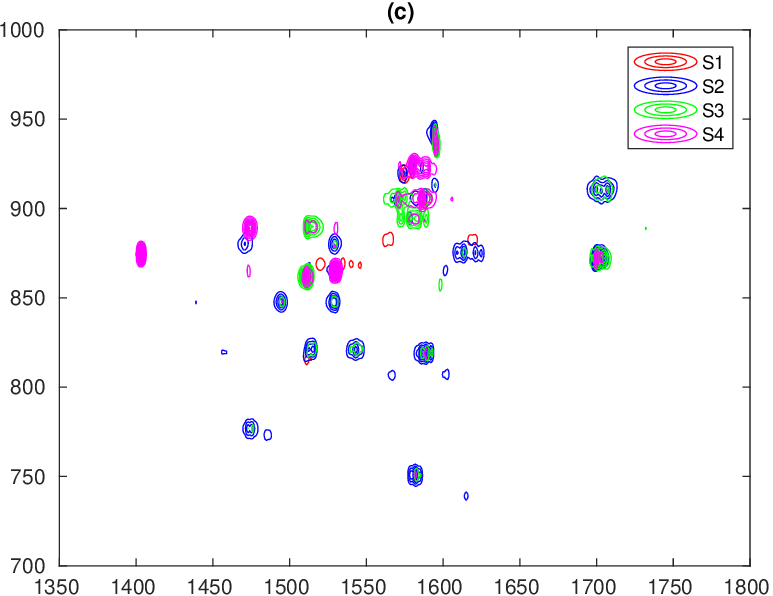}
\caption{\sar
{Contourplot of 2D HSCQ sources Limonene ($S_1$), Nerol ($S_2$), Terpinolene ($S_3$) and Caryophyllene ($S_4$). (a) pure sources, (b) (resp.(c)) estimated sources using the $\ell_1$ norm (resp. the $\beta$-divergence with the $\Ent$ regularization function).}
{2D HSQC sources (Limonene (red), Nerol (blue), Terpinolene (magenta) and Caryophyllene (green)): pure sources (a), zoom on the most important terpene zone 
using either the Frobenius norm with the $\ell_1$ norm (b), or the $\beta$-divergence with the $\Ent$ regularization function (c).}
}
\label{fig:2DNMRrslt}
\end{figure}

Table~\ref{Tab:Real_case} shows the real case with the optimal regularization parameter $\lambda_{\eS}$.
The $\beta$-divergence combined with $\ell_1$ norm or $\Ent$ function ensures the BSS of the 2D HSQC NMR data (see Fig.~\ref{fig:2DNMRrslt}).
However, compared with simulated data, we have a significant decrease of the SDR, SIR and SAR values which can probably  be explained by a wrong assumption on $\eD$ and possibly the linearity of the model.
This raises the question about the choice of the objective function $\Phi$ and requires further investigations to characterize the model in the 2D NMR context.

\begin{table}[htbp]
\centering 
\renewcommand{\arraystretch}{1.05}
\setlength{\tabcolsep}{0.1cm}
\begin{small}
\begin{tabular}{|c|c|c|c|c|c|} \hline
Data fidelity term & $\Psi_{\eS}$ & SDR &  SIR & SAR & Amari-index\\ \cline{1-6}
 
\multirow{2}{*}{Squared Frobenius} 
& $\ell_1 + \iota_+$ 
& 04.984 & 13.956 & 07.951 & 0.18037 \\ \cline{2-6}
& $\Ent + \iota_+$ 
& 05.755 & 14.434 & 08.446 & 0.17926 \\ \cline{1-6}

\multirow{2}{*}{$\beta$-divergence}
& $\ell_1 + \iota_+$ 
& \textbf{07.240} & \textbf{11.487} & 10.574 & 0.16098\\ \cline{2-6}

& $\Ent + \iota_+$ 
& 07.220 & 11.396 & \textbf{10.632} & \textbf{0.16526}\\ \cline{1-6}
\end{tabular}
\end{small}
\caption{Results on 2D real NMR data with $\lambda_{\eS}=10  \sigma$.}
\label{Tab:Real_case}
\end{table}

\newpage

\bibliographystyle{unsrt}
\bibliography{refs}

\end{document}